\documentclass[aps,twocolumn,prl,superscriptaddress]{revtex4-1}
\usepackage{graphicx}
\usepackage{epstopdf}
\usepackage{amsmath}
\usepackage{amsfonts}
\usepackage{amssymb}
\usepackage{makeidx}

\newcommand{\beq}{\begin{equation}} \newcommand{\eeq}{\end{equation}}
\newcommand{\bea}{\begin{eqnarray}} \newcommand{\eea}{\end{eqnarray}}
\newcommand{\bear}{\begin{eqnarray*}} \newcommand{\eear}{\end{eqnarray*}}
\newcommand{\lb}{\label}
\newcommand{\rf}[1]{(\ref{#1})}

\usepackage{hyperref}
\hypersetup{colorlinks=true, linkcolor=blue, citecolor=blue, filecolor=blue, urlcolor=blue,
            pdfproducer={Latex},
            pdfcreator={pdflatex}}

\begin{document}

\title {From an Action Principle for Action-dependent Lagrangians toward non-conservative Gravity: accelerating Universe without dark energy
 }


\author{Matheus J. Lazo}\email{matheuslazo@furg.br}
\author{Juilson Paiva}
\author{Jo\~ao T. S. Amaral}
\affiliation{Instituto de Matem\'{a}tica, Estat\'{\i}stica e F\'{\i}sica -- FURG, Rio Grande, RS, Brazil.}
\author{Gast\~ao S. F. Frederico}
\affiliation{Departamento de Matem\'{a}tica,
Universidade Federal de Santa Catarina, Florian\'opolis, SC, Brazil
}
\affiliation{Department of Science and Technology, University of Cape Verde, Praia, Cabo Verde.}

\begin{abstract}

In the present work, we propose an Action Principle for Action-dependent Lagrangians by generalizing the Herglotz variational problem for several independent variables. This Action Principle enables us to formulate Lagrangian densities for non-conservative fields. In special, from a Lagrangian depending linearly on the Action, we obtain a generalized Einstein's field equations for a non-conservative gravity and analyze some consequences of their solutions to cosmology and gravitational waves. We show that the non-conservative part of the field equations depends on a constant cosmological four-vector. Depending on this four-vector, the theory displays damped/amplified gravitational waves and an accelerating Universe without dark energy.



\end{abstract}

\maketitle


The Action Principle was introduced in its mature formulation by Euler, Hamilton
and Lagrange and, since then, it has become a fundamental principle for the construction of all physical theories.
In order to obtain the dynamical equations of any theory, the Lagrangian defining the Action is
constructed from the scalars of the theory. In this case, the action itself is a scalar.
Consequently, we might ask: what would happen if the Lagrangian itself is a function of the
Action? The answer to this question can be given by the Action Principle proposed by Herglotz
\cite{Herg1,Herg2,GGB}. The Herglotz variational calculus consists
in the problem of determining the path $x(t)$ that  extremize (minimize or maximize) $S(b)$,
where $S(t)$ is a solution of
\beq
\lb{H}
\begin{split}
\dot{S}(t)&=L(t,x(t),\dot{x}(t),S(t)),\;\; t\in [a,b]\\
S(a)&=s_a, \;\; x(a)=x_a, \;\; x(b)=x_b, \;\; s_a,x_a,x_b \in \mathbb{R}.
\end{split}
\eeq
It is easy to note that \rf{H} represents a family of differential equations since for each function $x(t)$ a different
differential equation arises. Therefore, $S(t)$ is a functional.
The problem reduces to the classical fundamental problem of the calculus of variations if the Lagrangian function
$L$ does not depend on $S(t)$. In this case we have $\dot{S}(t)=L(t,x(t),\dot{x}(t))$, and by integrating we obtain the classical variational problem
\beq
\lb{H2}
S(b)=\int_a^b \tilde{L}(t,x(t),\dot{x}(t))\;dt\longrightarrow {\mbox{extremum}},
\eeq
where $x(a)=x_a$, $x(b)=x_b$, and
\beq
\lb{H3}
\tilde{L}(t,x(t),\dot{x}(t))=L(t,x(t),\dot{x}(t)) +\frac{s_a}{b-a}.
\eeq
It is important to notice from \eqref{H2} that for a given fixed function $x(t)$ the functional $S$ reduces to a function of the domain boundary $a,b$. Herglotz proved \cite{Herg1,Herg2} that a necessary condition for a path $x(t)$ to be an extremizer of the variational
problem \eqref{H} is given by the generalized Euler-Lagrange equation:
\beq
\lb{HEL}
\frac{\partial L}{\partial x} -\frac{d}{dt}\frac{\partial L}{\partial \dot{x}}+\frac{\partial L}{\partial S}\frac{\partial L}{\partial \dot{x}}=0.
\eeq
In the simplest case where the dependence of the Lagrangian function on the Action is linear, the Lagrangian describes a dissipative system and, from \eqref{HEL}, the resulting equation of motion includes the well known dissipative term proportional to $\dot{x}$. It should also be noticed that in the case of the classical problem of the calculus of variation \eqref{H2} one has $\frac{\partial L}{\partial S}=0$, and the differential equation \eqref{HEL} reduces to the classical Euler-Lagrange equation.

In what follows we will be interested in a more general problem where the Lagrangian function depends on several independent variables $x^1,x^2,\cdots,x^d$ ($d=1,2,3,\cdots$). Besides, (as we are specially interested in the problem of gravity) we will consider a curved--space with metric $g_{\alpha \beta}=g_{\alpha \beta}(x^1,x^2,\cdots,x^d)$ defined on a domain $\Omega \subset \mathbb{R}^d$.
Thus, the classical problem of calculus of variation deals with the problem of finding $g_{\alpha \beta}$ that extremize the functional
\beq
\lb{H4}
S(\delta\Omega)=\int_{\Omega}\mathcal{L}(x^\mu,g_{\alpha \beta}, g_{\alpha \beta,\mu})\sqrt{•}\;d^d x,
\eeq
where, $g_{\alpha \beta,\mu}=\partial_\mu g_{\alpha \beta}$, $\sqrt{•}=\sqrt{|g|}$, $\delta\Omega$ is the boundary of $\Omega$, and $g_{\alpha \beta}$ satisfy the boundary condition $g_{\alpha \beta}(\delta\Omega)=g_{\alpha \beta}^{\delta\Omega}$ with $g_{\alpha \beta}^{\delta\Omega}:\delta \Omega \longrightarrow \mathbb{R}$. Unfortunately, despite the Herglotz problem was introduced in 1930, a covariant generalization of \eqref{H} for several independent variables is not direct and is lacking up to now. In order to generalize the Herglotz problem for fields, let us first note that, as in \eqref{H2}, for a given fixed $g_{\alpha \beta}$ the functional $S$ defined in \eqref{H4} reduces to a function of the boundary $\delta \Omega$.
Let now we consider that $\delta\Omega$ is an orientable Jordan surface with normal $n^\mu$. If there is a differentiable vector field $s^\mu$ such that
\beq
\lb{H5}
S(\delta\Omega)=\int_{\delta\Omega}n_\nu s^\nu \sqrt{|h|}\; d^{d-1}x,
\eeq
where $\sqrt{|h|}$ is the induced metric over $\delta\Omega$, then we obtain
\beq
\lb{H6}
\begin{split}
S(\delta\Omega)&=\int_{\delta\Omega}n_\nu s^\nu \sqrt{|h|} \; d^{d-1}x=\int_{\Omega} \nabla_\nu s^\nu \sqrt{•}\; d^dx\\
&=\int_{\Omega}\mathcal{L}(x^\mu,g_{\alpha \beta},g_{\alpha \beta,\mu}) \sqrt{•}\;d^dx,
\end{split}
\eeq
where we used the Stokes' Theorem and $\nabla_\nu$ stands for a covariant derivative. Consequently, we can generalize the Action Principle by stating that the space-time metric $g_{\mu\nu}$ is that which extremize the action $S(\delta\Omega)$ given by
\beq
\lb{H7}
\begin{split}
\nabla_\nu s^\nu &=\mathcal{L}(x^\mu,g_{\alpha \beta},g_{\alpha \beta,\mu},s^\mu),\;\;\; x^\mu \in \Omega\\
S(\delta\Omega)&=\int_{\delta\Omega}n_\nu s^\nu \sqrt{|h|}\; d^{d-1}x, \;  g_{\alpha \beta}(\delta\Omega)=g_{\alpha \beta}^{\delta\Omega},
\end{split}
\eeq
where $g_{\alpha \beta}^{\delta\Omega}$ is fixed. It is important to notice that our Action Principle \eqref{H7} (that generalizes \eqref{H} for fields) reduces to the classical Action Principle if the Lagrangian is independent on $s^\mu$. Furthermore, for the case where $s^\nu=(s^0,0,0,0)$ and $\Omega=[t_a,t_b]\otimes \mathbb{R}^3$, \eqref{H7} contain as a particular case the non-covariant problem introduced in \cite{GGB}. Moreover, in this last situation \eqref{H7} can be easily solved for Lagrangians linear on $s^0$, giving a $s^0$ expressed as a history-dependent function of the source.


For the gravity field, the Lagrangian we propose is given by $\mathcal{L}=\mathcal{L}_m+\mathcal{L}_g$, where $\mathcal{L}_m$ is the Lagrangian for matter and
\beq
\lb{H8}
\mathcal{L}_g(x^\mu,g_{\alpha \beta},g_{\alpha \beta,\mu},s^\mu) = R-\lambda_\nu s^\nu,
\eeq
where $\lambda_\nu$ is a constant cosmological four-vector. In \eqref{H8}, $R=\tilde{L}-L$ is the Ricci scalar with $\tilde{L}=g^{\mu\nu}(\Gamma^{\sigma}_{\mu\sigma,\nu}-\Gamma^{\sigma}_{\mu\nu,\sigma})$ and $L=g^{\mu\nu}(\Gamma^{\sigma}_{\mu\nu}\Gamma^{\rho}_{\sigma\rho}-\Gamma^{\rho}_{\mu\sigma}\Gamma^{\sigma}_{\nu\rho})$.
Since the second order derivatives in \eqref{H8} occur only linearly in the Lagrangian, the field equations
can be obtained by an effective Lagrangian $\mathcal{L}=\mathcal{L}_m+\mathcal{L}_{ef}$ with
\beq
\lb{H9}
\mathcal{L}_{ef}(x^\mu,g_{\alpha \beta},g_{\alpha \beta,\mu},s^\mu) = L-\lambda_\nu s^\nu,
\eeq
instead of \eqref{H8}, because $\int_{\Omega}\;\tilde{L} \sqrt{•} \; d^{d}x=2\int_{\Omega}\;L \sqrt{•}\; d^{d}x+\mbox{constant}$ (see \cite{Dirac}).


In order to obtain the generalized field equations, let us define a family of metrics $g_{\alpha \beta}$ such that
\beq
\lb{PH1}
g_{\alpha \beta}(x^\mu)=g_{\alpha \beta}^*(x^\mu)+\delta_\epsilon (g_{\alpha \beta})(x^\mu),
\eeq
where $g_{\alpha \beta}^*$ is the metric that extremize $S(\delta\Omega)$ in \eqref{H7}, $\epsilon \in \mathbb{R}$, and $\delta_\epsilon (g_{\alpha \beta})$ satisfies the boundary condition $\delta_\epsilon (g_{\alpha \beta})(\delta\Omega)=0$ and $\lim_{\epsilon\rightarrow 0}\delta_\epsilon (g_{\alpha \beta})(x^\mu)=0$
(weak variations).
Since $S(\delta\Omega)$ attain a extremum at $g_{\alpha \beta}^*$, we have
\beq
\lb{PH2}
\lim_{\epsilon\rightarrow 0}\frac{\delta_{\epsilon} (S)(\delta\Omega)}{\epsilon}=0.
\eeq
From \rf{H5} we get
\beq
\lb{PH3}
\lim_{\epsilon\rightarrow 0}\frac{\delta_{\epsilon} (S)(\delta\Omega)}{\epsilon}=\int_{\delta\Omega}n_\nu \lim_{\epsilon\rightarrow 0}\frac{\delta_{\epsilon} (s^\nu)}{\epsilon}\sqrt{|h|} \; d^{d-1}x =0,
\eeq
since the surface $\delta\Omega$, and consequently $\sqrt{|h|}$, is independent on $\epsilon$. A sufficient condition to satisfy \eqref{PH3} for arbitrary boundary $\delta\Omega$ is that
\beq
\lb{PH3b}
\lim_{\epsilon\rightarrow 0}\frac{\delta_{\epsilon} (s^\nu)(\delta\Omega)}{\epsilon}=0.
\eeq
On the other hand, by integrating over $\Omega$ both sides of the differential equation in \eqref{H7} we obtain
\beq
\lb{PH5}
S(\delta\Omega)=\int_{\Omega}\mathcal{L}(x^\mu,g_{\alpha \beta},g_{\alpha \beta,\mu},s^\mu) \sqrt{•}\;d^dx,
\eeq
and by taking the variation of \eqref{PH5} we get
\beq
\lb{PH6}
\begin{split}
&\delta_{\epsilon} (S)=\int_{\Omega} \delta_{\epsilon}(\mathcal{L}(x^\mu,g_{\alpha \beta},g_{\alpha \beta,\mu},s^\mu)\sqrt{•})\;d^dx\\
&\;\;\;\;\;=\int_{\Omega}\left[ \delta_{\epsilon}(L\sqrt{•}) + \delta_{\epsilon}(\mathcal{L}_m\sqrt{•})-\lambda_\nu\delta_{\epsilon} (s^\nu\sqrt{•})\right]\;d^dx.
\end{split}
\eeq
We also have from \eqref{H6}, by using $\nabla_\nu (\cdot) \sqrt{•}=\partial_\nu (\cdot \sqrt{•})$,
\beq
\lb{PH7}
\delta_{\epsilon} (S)= \delta_{\epsilon}\int_{\Omega} \nabla_\nu s^\nu \sqrt{•}\; d^dx=\int_{\Omega} \partial_\nu \delta_{\epsilon}(s^\nu \sqrt{•})\; d^dx.
\eeq
From \eqref{PH6} and \eqref{PH7} we obtain
\beq
\lb{PH8}
\int_{\Omega} \left[\partial_\nu \delta_{\epsilon}(s^\nu \sqrt{•}) -\delta_{\epsilon}((L+\mathcal{L}_m)\sqrt{•})+\lambda_\nu\delta_{\epsilon} (s^\nu\sqrt{•})\right]\; d^dx=0.
\eeq
Since \eqref{PH8} should be satisfied for any domain $\Omega$, we have
\beq
\lb{PH9}
\partial_\nu \zeta^\nu  =\delta_{\epsilon}(L\sqrt{•})+\delta_{\epsilon}(\mathcal{L}_m\sqrt{•}) -\lambda_\nu\zeta^\nu,
\eeq
where $\zeta^\nu=\delta_{\epsilon}(s^\nu\sqrt{•})$. Due to $\lambda_\nu$ be a constant four-vector, \eqref{PH9} implies that $\zeta^\nu$ can be written as
\beq
\lb{PH10}
\zeta^\nu(\epsilon)=A^\nu(x^\mu,g_{\alpha \beta},g_{\alpha \beta,\mu},s^\mu)e^{-\lambda_\gamma x^\gamma},
\eeq
where
\beq
\lb{PH11}
\partial_\nu A^\nu=\left(\delta_{\epsilon}(L\sqrt{•})+\delta_{\epsilon}(\mathcal{L}_m\sqrt{•})\right)e^{\lambda_\gamma x^\gamma}.
\eeq
From \eqref{PH3b} we should have, since $\delta_\epsilon (g_{\mu\nu})(\delta\Omega)=0$,
\beq
\lb{PH12a}
\zeta^\nu (0)=A^\nu|_{\epsilon=0} e^{-\lambda_\gamma x^\gamma}=0
\eeq
for all $x^\mu \in \delta\Omega$. As a consequence, $A^\nu$ is identically zero over $\delta\Omega$. In this case, we obtain from Stokes' Theorem
\beq
\lb{PH12}
\int_{\delta\Omega}
n_\nu \frac{A^\nu}{\sqrt{•}}  \sqrt{|h|}\; d^{d-1}x=\int_{\Omega} \partial_\nu A^\nu \; d^dx=0.
\eeq
Thus
\beq
\lb{PH13}
\begin{split}
&\int_{\Omega}\delta_{\epsilon}(L\sqrt{•}+\mathcal{L}_m\sqrt{•})e^{\lambda_\gamma x^\gamma}\;d^dx\\
&=\int_{\Omega}\left[\Gamma^\alpha_{\mu\nu}\delta_{\epsilon} (g^{\mu\nu}\sqrt{•})_{,\alpha}-\Gamma^\alpha_{\mu\alpha}\delta_{\epsilon} (g^{\mu\nu}\sqrt{•})_{,\nu}\right.\\
&\qquad \quad + \left(\Gamma^\beta_{\mu\alpha}\Gamma^\alpha_{\nu\beta}-\Gamma^\beta_{\alpha\beta}\Gamma^\alpha_{\mu\nu} \right)\delta_{\epsilon} (g^{\mu\nu}\sqrt{•})\\
&\left. \qquad \quad +8\pi G T_{\mu\nu} \delta_{\epsilon}(g^{\mu\nu})\sqrt{•}\right]e^{\lambda_\gamma x^\gamma}\;d^dx\\
&=-\int_{\Omega}\delta_{\epsilon} (g^{\mu\nu})\sqrt{•} \left[R_{\mu\nu}-\frac{1}{2}g_{\mu\nu}R+K_{\mu\nu}\right. \\
&\qquad \quad \left. -\frac{1}{2}g_{\mu\nu}K-8\pi G T_{\mu\nu} \right]e^{\lambda_\gamma x^\gamma}\;d^dx\\
&\quad +\int_{\Omega} \Big[ \Big(\Gamma^\alpha_{\mu\nu}\delta_{\epsilon} (g^{\mu\nu}\sqrt{•})\\
&\qquad \quad -\Gamma^\nu_{\mu\nu}\delta_{\epsilon} (g^{\mu\alpha}\sqrt{•})\Big)e^{\lambda_\gamma x^\gamma} \Big]_{,\alpha}\;d^dx=0,
\end{split}
\eeq
where we define the symmetric tensor $K_{\mu\nu}=\lambda_{\alpha}\Gamma^\alpha_{\mu\nu}-\frac{1}{2}\left(\lambda_{\nu}\Gamma^\alpha_{\mu\alpha}+\lambda_{\mu}\Gamma^\alpha_{\nu\alpha}\right)$, and $\delta_{\epsilon}(\mathcal{L}_m\sqrt{•})=\frac{8\pi G}{c^4} T_{\mu\nu}\delta_{\epsilon}(g^{\mu\nu})\sqrt{•}$ where $T_{\mu\nu}$ is the energy-momentum tensor.
The last integral in \eqref{PH13} is zero since $\delta_\epsilon (g_{\mu\nu})(\delta\Omega)=0$. Thus, from the Fundamental Lemma of calculus of variation we obtain from \eqref{PH13} the generalized gravitational field equation
\beq
\lb{GEEQ}
R_{\mu\nu}+K_{\mu\nu}-\frac{1}{2}g_{\mu\nu}\left(R+K\right)=\frac{8\pi G}{c^4} T_{\mu\nu}.
\eeq

It is important to remark that the generalized gravity field \eqref{GEEQ} depending on the cosmological four-vector $\lambda_\mu$
can be used to describe non-conservative phenomena, since the covariant divergent $\nabla_\mu(K^\mu_{\nu}-\frac{1}{2}g^\mu_\nu K)$ is in general different from zero for $\lambda_\mu\neq 0$.
A notable consequence from this non-conservation is that the space-time manifold behaves just similar to a non-perfectly elastic rubber sheet.
In order to shed light on the effects of the non-conservation on the geometrical side of the field equation \eqref{GEEQ} when $\lambda_{\mu} \neq 0$, it is interesting to investigate the behavior of gravitational waves. We suppose the metric to be close to the Minkowski one \cite{weinberg}, i.e.
$g_{\mu\nu}=\eta_{\mu\nu}+h_{\mu\nu}$ with $|h_{\mu\nu}|\ll 1$.
To first order in $h$, by choosing the modified harmonic gauge $\eta^{\mu \nu}(h_{\mu \rho,\nu}-\frac{1}{2}h_{\mu \nu,\rho}+\lambda_\mu h_{\nu \rho}-\frac{1}{2}\lambda_\rho h_{\mu \nu})=0$, we
obtain from the field equations
\begin{equation}\label{eq:field-eq}
\begin{split}
\square^2 &h_{\mu\nu}+\lambda^{\rho} h_{\mu\nu,\rho} =-\frac{16\pi G}{c^4}S_{\mu\nu}
\end{split}
\end{equation}
where $S_{\mu\nu}=T_{\mu\nu}-\frac 1 2\eta_{\mu\nu}T^\lambda_{\;\;\lambda}$.
For simplicity, let us consider only the homogeneous case $S_{\mu\nu}=0$ with $\lambda_1=\lambda_2=\lambda_3=0$, and a gravitational wave traveling on the $x^3=z$ direction. In this case $h_{\mu \nu}$ is a function on $t$ and $z$ and we also have $h_{0\mu}=h_{3\mu}=0$. From the wave equation \eqref{eq:field-eq}
we obtain three possible solutions for $h_{\mu \nu}$:
\begin{equation}
h_{\mu \nu}(t,z)=
\left\{
\begin{array}{ll}
h_{\mu \nu}^{(\pm)}e^{-\frac{\lambda_0 \pm \lambda^\prime}{2}ct}e^{ikz} & \mbox{if $\lambda_0^2>4k^2$};\\
\left( h_{\mu \nu}^{(+)}+h_{\mu \nu}^{(-)}ct \right) e^{-\frac{\lambda_0}{2}ct}e^{ikz} & \mbox{if $\lambda_0^2=4k^2$};\\
h_{\mu \nu}^{(\pm)}e^{-\frac{\lambda_0 \pm i\lambda^\prime}{2}ct}e^{ikz} & \mbox{if $\lambda_0^2<4k^2$},\\
\end{array}
\right. \nonumber
\end{equation}
where $\lambda^\prime\equiv\sqrt{|\lambda_0^2-4k^2|}$, and $h_{\mu \nu}^{(\pm)}$ are constant symmetric tensors with non-null components $h_{1 1}^{(\pm)}$, $h_{2 2}^{(\pm)}=-h_{1 1}^{(\pm)}$ and $h_{1 2}^{(\pm)}$. When $\lambda_0>0$ ($\lambda_0<0$)
we observe three cases of damped (amplified) waves and, in any of these cases, the amplitude of
gravitational waves decreases (increases) with time. It is important to notice that both $\lambda_0^2>4k^2$ and $\lambda_0^2=4k^2$
solutions corresponds to stationary waves and occur for small spatial frequencies ($k \le |\lambda_0|/2$). On the other hand, the solution when $\lambda_0^2<4k^2$ corresponds to traveling waves with velocity $v=\frac{\lambda^\prime}{2 k}c$, smaller than the speed of light $c$. Furthermore, the dispersion relation $\omega=\frac{\lambda^\prime}{2}c$ relating time and space frequencies give us an experimental test for the existence of the cosmological four-vector $\lambda_\mu$.

Despite the non-conservation on the geometrical side of the field equation \eqref{GEEQ}, there are two simple possibilities in order to enable solutions where we have energy-momentum conservation (that implies $T^\mu_{\nu;\mu}=0$). The first is to change how mass-energy generates curvature by considering that the gravity constant $G$ is actually a function on $x^\mu$. In this approach, the function $G$ equalize the conserved matter side in \eqref{GEEQ} with a non-conservative geometry. The second possibility is to introduce a cosmological constant $\Lambda$ in the theory that is actually a function on $x^\mu$. The cosmological constant can easily be included by adding $-2\Lambda$ in the Lagrangians \eqref{H8} and \eqref{H9}. For simplicity, in the present work we consider only the first case where, by taking the covariant derivative of \eqref{GEEQ} with $\nabla_\mu T^\mu_{\nu}=0$, we have the conservation condition
\beq
\lb{CC}
\nabla_\mu\left(K^\mu_{\nu}-\frac{1}{2}g^\mu_\nu K\right)=8\pi G_{,\mu} T^\mu_{\nu}.
\eeq

\begin{figure}[ht]
\includegraphics[width=8.0cm]{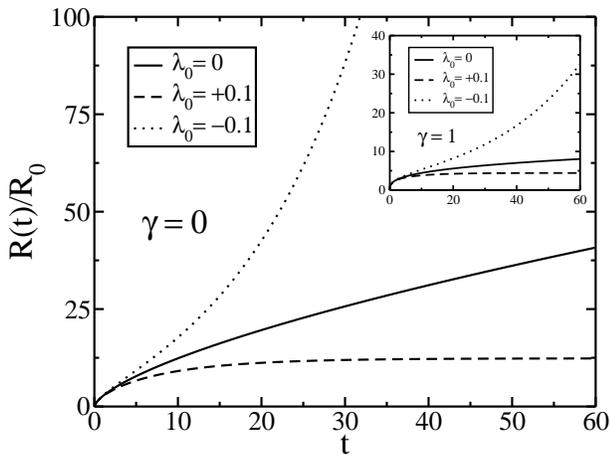}
\caption{The figure displays the isotropic scale factor $R(t)$ versus $t$ (for a cosmological time scale) in a matter-dominated era ($\gamma=0$), with $G_0=1$, $p=\gamma \rho $, and where $R_0$ is a constant. The inset shows the strong radiation-dominated era ($\gamma=1$).
}
\label{fig}
\end{figure}

Finally, in order to investigate the cosmological consequences of the constant four-vector  $\lambda_\mu$, we analyze the dynamics of a Bianchi $I$ universe filled with a perfect fluid. The metric we consider is given by \cite{Bianchi}
\beq
\lb{metric}
ds^2=dt^2 - a_1^2(t) dx^2 - a_2^2(t) dy^2 - a_3^2(t) dz^2,
\eeq
where we set $c=1$ for simplicity. From the field equation \eqref{GEEQ} and from \eqref{CC} we get
\beq
\lb{BME}
\begin{split}
\frac{\dot{a}_1}{a_1}\frac{\dot{a}_2}{a_2}+\frac{\dot{a}_1}{a_1}\frac{\dot{a}_3}{a_3}+\frac{\dot{a}_2}{a_2}\frac{\dot{a}_3}{a_3}=8\pi G\rho &=-\frac{4\pi}{\lambda_0} \dot{G}\rho \\
\frac{\ddot{a}_i}{a_i}+\frac{\ddot{a}_j}{a_j}+\frac{\dot{a}_i}{a_i}\frac{\dot{a}_j}{a_j}+\lambda_0\left(\frac{\dot{a}_i}{a_i}+\frac{\dot{a}_j}{a_j}\right)&=-8\pi Gp, \; i\neq j.
\end{split}
\eeq
where we consider $\lambda_1=\lambda_2=\lambda_3=0$, and $T_{\mu\nu}= (\rho +p)U_\mu U_\nu - pg_{\mu\nu}$ for the perfect fluid (where $\rho$ is the matter density, $p$ is the pressure and $U_\mu$ is the fluid velocity), with pressure $p$ and density $\rho$
obeying the equation of state $p=\gamma\rho$ $(0\leq \gamma \leq 1)$ \cite{Carroll}. From the first equation in \eqref{BME} we obtain
\beq
\lb{GC}
G(t)=G_0e^{-2\lambda_0 t},
\eeq
where $G_0$ is a constant. It is important to notice from \eqref{GC} that for $\lambda_0 < 0$ ($\lambda_0 > 0$) the coupling $G$ between geometry and matter is strengthened (weakened) as a consequence of the non-conservation in the geometrical side of the field equation \eqref{GEEQ}. In Figure \ref{fig} we display the isotropic solution of the scale factor $a_1(t)=a_2(t)=a_3(t)=R(t)$, in the cases
where $\gamma$ takes the values $0$ and $1$, respectively corresponding to matter and strong radiation-dominated eras. In both cases,
one can see that the most important consequence of the constant cosmological four-vector is the arising of a Universe with
accelerated expansion rate when $\lambda_0<0$ without the necessity of introducing dark energy. The accelerated expansion rate is evident from the concavity inversion for $R(t)$ when $\lambda_0<0$.
For an isotropic matter-dominated era this concavity inversion occurs at a time $t^*=\frac{1}{|c\lambda_0 |}\ln\left(\frac{3}{2}\right)$. Consequently, from observational evidence \cite{FTH} we should have $|\lambda_0|c$ of order $10^{-10}$yr$^{-1}$. Despite we consider a very simple Bianchi $I$ cosmological model, this result is in good agreement with observational and experimental bounds to the temporal rate of variation for $G$ \cite{Uzan}.
Furthermore, although we displayed in Figure \ref{fig} only the isotropic case, we have checked that the same behavior is obtained in the more general anisotropic case.
Actually, we expect that the same phenomenon will be present in more realistic models since the main mechanism behind the accelerated expansion is the non-conservation in the geometrical side of the field equation \eqref{GEEQ}. Finally, from Figure \ref{fig} it is also evident that when $\lambda_0>0$ the Universe reaches quickly a stationary state. Furthermore, in this case, the weakening of the coupling \eqref{GC} for $\lambda_0>0$ results in the asymptotic decoupling between matter and geometry.

Lastly, due to its smallness, the effects of $\lambda_0$ in the solar system for non-cosmological time scales is very small. For a short time interval, it is easy to verify from \eqref{eq:field-eq} that a spherically symmetric mass distribution reproduces the Newtonian gravity for weak fields since, in this case, we get $h_{00}= \frac{2\phi}{c^ 2}$, where $\phi$ is the Newtonian gravitational potential. Furthermore, as the metric should be a smooth function of time, we can estimate an upper limit of only $|\Delta \theta_{\lambda}-\Delta \theta_{0}|\lesssim 10^{-7}$ seconds of arc per century for the difference between the Mercury precession $\Delta \theta_{\lambda}$ in our theory (with $|\lambda_0|c\approx 10^{-10}$) and the precession of $\Delta \theta_{0} = 43.03''$ per century in classical gravity \cite{weinberg}.



In conclusion, in this work, we presented a generalization of the Action Principle for Action-dependent Lagrangians and considered it on a curved space with metric $g_{\mu\nu}(x^\mu)$. From this Action Principle, we obtained a generalized gravitational field equation, which can be used in the description of non-conservative phenomena. An interesting feature of this theory is that the gravitational field depends on a constant cosmological four-vector. The potential importance of this new gravitational theory is evident when applied to the problem of gravitational waves and to cosmology. Depending on the cosmological four-vector, we show that gravitational waves propagate with velocity smaller than the speed of the light, and with amplitudes which decrease (or increase) with time. Moreover, application to cosmology led to another remarkable result: a Universe (here considered as filled with a perfect fluid) displaying an accelerated expansion rate with no need to introduce dark energy. Finally, there are many directions of investigation left to explore related to developments of our former results. In special  we outline the post-Newtonian limit for a spherically symmetric mass distribution, enabling the investigation of the stability of planetary orbits in cosmological time scales, and the effects on galaxy rotations. Furthermore, although we consider only the gravitational problem, the Action Principle we propose is general and can be easily extended to any physical field.


\begin{acknowledgments}
M. J. Lazo would like to dedicate this work in memory of professor Silvestre Ragusa. This work was partially supported by CNPq and CAPES (Brazilian research funding agencies).
\end{acknowledgments}


\end{document}